# Gossip Codes for Fingerprinting: Construction, Erasure Analysis and Pirate Tracing

**Ravi S. Veerubhotla**
(Institute for Development and Research in Banking Technology, Hyderabad, India
vrsankar@idrbt.ac.in)

**Ashutosh Saxena**
(Institute for Development and Research in Banking Technology, Hyderabad, India
asaxena@idrbt.ac.in)

**V.P. Gulati**
(Institute for Development and Research in Banking Technology, Hyderabad, India
vpgulati@idrbt.ac.in)

**A.K. Pujari**
(Department of Computer & Information Sciences, University of Hyderabad, India
akpcs@uohyd.ernet.in)

**Abstract:** This work presents two new construction techniques for $q$-ary Gossip codes from $t$-designs and Traceability schemes. These Gossip codes achieve the shortest code length specified in terms of code parameters and can withstand erasures in digital fingerprinting applications. This work presents the construction of embedded Gossip codes for extending an existing Gossip code into a bigger code. It discusses the construction of concatenated codes and realisation of erasure model through concatenated codes.

**Categories:** H.1.1, H.5.1, H.3.7

## 1 Introduction

Protecting digital content from illegal copying and distributing is one of the key issues worrying owners and distributors in the digital world. Copyrights are an accepted means of legally encouraging the creation of works and their rightful exploitation. Recent developments in digital information processing and distribution have had a tremendous impact on the creators and their copyrights. On the one hand, these advancements open new ways of creating and exploiting digital works, while on the other, the same advancements also open up novel ways of circumventing copyrights. These techniques can be used against the interests of copyright holders by creating less number of copies and distributing them effectively. Therefore, new digital rights protection techniques are required, which allow the creators and other right-holders to enforce their legal rights and interests. The overall goal of these techniques is to ensure sufficient incentives for creating works and making them available in the digital form.

Fingerprinting digital objects help in checking copyright violations in the digital world. The scope of digital fingerprinting also includes pay per view, pay per use and other broadcast applications. Digital fingerprinting refers to the act of embedding a unique identifier in a digital object, in such a way that it is difficult for others to find and destroy the identifier. The fingerprints are typically embedded into the content using watermarking techniques, which are designed to be robust to a variety of attacks. This marking makes every user's copy unique, while still being close to the original. It allows the distributor to detect any unauthorised copy and trace it back to its origin. Since a marked object can be traced back to the owner, users will be deterred from releasing an unauthorised copy. A cost-effective attack against such digital fingerprints is collusion, where several differently marked copies of the same content are combined to disrupt the underlying fingerprints.

A coalition of users may detect the marks by comparing multiple fingerprinted objects and modify or erase them. In this process, pirates may further try to frame innocent users. To prevent this, Boneh and Shaw [Boneh and Shaw 98] introduced frameproof codes in which the pirates cannot frame innocent users outside their coalition. Moreover, it is necessary for the distributor to find at least one user involved in creating a pirate copy. Identifiable Parent Property (IPP) codes, introduced by Hollman et al. [Hollman, van Lint, Linnartz and Tolhui-zen 98] exactly for this purpose, can identify at least one parent of the pirate word. Non-binary (alphabet size $q > 2$) IPP codes can handle a large collusion size $c$ ($> 2$) [Barg, Cohen, Encheva, Kabatiansky and Zémor 01] unlike binary IPP codes. In [Trappe, Wu, Wang and Liu 03], the authors investigate the problem of designing fingerprints that can withstand collusion attacks and allow the identification of colluders using Anti-Collusion codes (ACC), which were originally proposed in [Trappe, Wu, Wang and Liu 02]. Their work addresses the collusion problem considering the additive embedding and then study the effect that an averaging collusion has upon orthogonal modulation. Their fingerprinting scheme is based upon code modulation and the code construction requires only $O(\sqrt{M})$ orthogonal signals in order to accommodate $M$ users, for a given number of colluders. The work presents four different detection strategies that can be used with ACC for identifying a suspect set of colluders.

Traceable schemes are cryptographic systems that provide protection against illegal copying and redistribution of digital data. A closely related concept is traceability systems introduced by Chor et al [Chor, Fiat, Naor 94] used in the context of broadcast encryption schemes. Broadcast encryption systems [Fiat, Naor 93], [Stinson, Wei 99] allow targeting of an encrypted message to a privileged group of receivers. Each receiver has a decoder with a set of keys that allows him to decrypt the encrypted messages if he is a member of the target group. When a group of colluders (up to $c$) construct a pirate decoder to decrypt the content, broadcast encryption systems can identify at least one of the colluders who contributed to the pirate decoder. In the tracing mechanism proposed in [Stinson and Wei 98], the merchant finds the numbers of common keys between the pirate decoder and all the authorised decoders. The merchant then chooses the decoder with the maximum number of common keys with the pirate decoder, which exposes a culprit. These schemes are known as Traceability schemes and are also called *key fingerprinting schemes,* which allow tracing of one of the colluders. In a Traceability Scheme [Chor,

Fiat, Naor 94], [Stinson and Wei 98], [Kurosawa, Desmedt 98], [Kurosawa, Desmedt 98],[Garay, Staddon, Wool 99], each authorised user has a decoder with a set of $k$ keys, from a base key set $X$ of size $v$ that uniquely determines the owner and allows him to decrypt the encrypted broadcast.

This work present the construction methods for non-binary IPP codes, namely, $c$- Gossip codes [Lindkvist 01] from $t$-designs, Traceability schemes and vice versa. The constructions attain the minimum possible code length specified for Gossip codes, in terms of alphabet size $q$, number of codewords $M$ and collusion size $c$. We describe the construction of embedded Gossip codes and also generalise Gossip codes. We discuss the pirate tracing mechanism for these codes in presence of erasures. We also present a construction method and an analysis of Concatenated codes.

The organization of the paper is as follows: Section 2 presents the preliminaries, namely Traceability schemes, $t$-designs, fingerprinting codes and related work. In section 3, we describe the construction of $c$- Gossip codes with shortest possible length from $c$-Traceability schemes, $t$-designs and vice versa. This section also presents the construction of embedded Gossip codes and generalised Gossip codes. The section 4 models erasures in Gossip codes, section 5 analyses Gossip codes and concatenated codes followed by the conclusions in Section 6.

## 2   Preliminaries

This section presents the background and related work on fingerprinting.

**Definition 2.1** Consider a code $\Gamma$ of length $N$ on an alphabet $Q$ with $|Q|=q$ whose symbols are denoted by $\{0, ..., q-1\}$. Then $\Gamma \subseteq Q^N$ and we will call it an $(N, n, q)$-code if $|\Gamma|=n$. The elements of $\Gamma$ are called codewords. Each codeword is given by $x = (x_1, ..., x_N)$, where $x_i \in Q$, $1 \le i \le N$. For any subset of codewords $C \subseteq \Gamma$, we define the set of descendents of $C$, denoted $D(C)$ by $D(C) = \{x \in Q^N : x_i \in \{a_i : a \in C\}, 1 \le i \le N\}$. (see [Stinson and Wei 98])

The set $D(C)$ consists of $N$-tuples that could be produced by a coalition holding the codewords of the set $C$.

### 2.1   Gossip codes

Gossip codes are fingerprinting codes introduced in [Lindkvist 01] which can provide protection against illegal copying of digital objects. They are also $c$-IPP codes, which can identify at least one user involved in creating an illegal copy.

We first explain the construction of Gossip codes originally presented in [Lindkvist, Löfvenberg, and M. Svanström 02].
Let $B(M, q)$ be the $0/1$-matrix consisting of $l$ columns and $M$ rows such that each column is created by placing $q-1$ ones and $M-q+1$ zeros. The parameters $q$ and $M$ are so chosen satisfying $q \ge 3$ and $M \ge q+1$. The codeword matrix $G(M, q)$ is constructed from $B(M, q)$ by replacing the $q-1$ ones in each column, with the $q-1$ different non-zero symbols of $Q$ and retaining the zeros unaltered. The code matrix $G(M, q)$ is called a Gossip code and each code column is known as a gossip column.

In a gossip column, the occurrence of non-zero alphabet symbols (among themselves) is immaterial as long as they are distinct. We denote $G(M, q)$ as $c\text{-}Gossip(l, M, q)$ code where $c$ is collusion size, $M$ is the number of code words, $q$ is the alphabet size and $l$ is the length of the code. With the above construction method a 2-Gossip (7, 7, 4) code constructed is as follows.

Example 2.1.1    2-Gossip (7, 7, 4) code

| 1 | 1 | 1 | 0 | 0 | 0 | 0 |
|---|---|---|---|---|---|---|
| 2 | 0 | 0 | 1 | 1 | 0 | 0 |
| 3 | 0 | 0 | 0 | 0 | 1 | 1 |
| 0 | 2 | 0 | 2 | 0 | 0 | 2 |
| 0 | 3 | 0 | 0 | 2 | 2 | 0 |
| 0 | 0 | 2 | 3 | 0 | 3 | 0 |
| 0 | 0 | 3 | 0 | 3 | 0 | 3 |

**Definition 2.1.1**  A gossip column is a code column of a $q$-ary code (Gossip code), with all non-zero symbols of the alphabet $Q$ appearing exactly once, and the symbol '*zero*' in the remaining positions. (see [Lindkvist 01])

### 2.1.1    Tracing using gossip columns

Consider the 6[th] column of the Gossip code presented in e.g. 2.1.1. In this gossip column the non-zero symbols are distinct and appear at positions 3[rd], 5[th] and 6[th]. Thus this gossip column is responsible for tracing users {3[rd] or 5[th] or 6[th]} (based on 6[th] position in the pirate word) if they are involved as traitors. The collusion groups whose members can be traced (accused) by a gossip column are known as accusation groups of that column. Thus this gossip column accuse the collusion groups {3, 5}, {5, 6} and {3, 6} of collusion size $c = 2$. The remaining collusion groups are accusation groups for some other columns. The collusion groups {3, $i$}(for $i$=1, 2, 4 or 7) certainly can create a pirate word with a '*one*' at that column's position. But the tracing of these groups is left over to other gossip columns. Thus a gossip column contributes to tracing traitors involved in creating a pirate word.

**Lemma 2.1.1** A Gossip code is the code with a gossip column for each collusion group to trace. (see [Lindkvist 01])

The tracing method is based on the fact that every illegal fingerprint contains at least one non-zero symbol. If a particular collusion group is traceable by a gossip column, then all its subgroups are also traceable by the same gossip column since they will have a subset of possible alphabet symbols. If there exists a gossip column for each $c$-group, it is possible to trace at least one member of all pirate groups whose size is less than or equal to *c*. So, the Gossip code will always find a member of the pirate group and thus it is a *c*-IPP code.

**Definition 2.1.2** The set of non-zero positions of a gossip column is termed as column key corresponding to that column.

In e.g. 2.1.1 the column key for the 6$^{th}$ column is {3, 5, 6}. *Column keys* of Gossip code identify the set of pirate groups that each gossip column is capable of accusing. From a gossip column of a Gossip code, where $c < q < M$, exactly $\binom{q-1}{c}$ collusion groups are traceable. If $M$ is the total number of codewords in a Gossip code, then the total number of possible pirate groups of size *c* is given by $\binom{M}{c}$. For $c \leq q-1$, each gossip column at the most can accuse $\binom{q-1}{c}$ groups since there are $(q-1)$ non-zero symbols, which are distinct. Hence, in a Gossip code the total number of gossip columns $l \geq \binom{M}{c} / \binom{q-1}{c}$ for $c < q < M$. This gives a lower bound on length of the Gossip code. If the Gossip code length attains the equality condition it has the shortest possible length. Hereafter we denote this shortest length by $l$.

The structure of Gossip code is such that every non-zero symbol in the pirate word will lead to one culprit. (see Lemma 4.1.1)  Since Gossip codes are *q*'ary IPP codes and come up with a deterministic tracing algorithm unlike *q*'ary *c*-secure code with error probability $\varepsilon$ [Boneh and Shaw 98], they have a significant role to play in fingerprinting applications.

**2.2    Traceability Schemes**

A *c*-Traceability scheme *c-TS(k, b, v)* [Stinson,Wei 99], [Stinson and Wei 98] is a broadcast encryption scheme, where *k* is the number of keys provided to each user, *b* is the total number of users, *v* is the total number of base keys and *c* is the collusion size of pirates that the scheme can withstand.  The Trusted Authority generates a set *T* of *v* base keys and assigns *k* keys chosen from *T* to each user. The $i^{th}$ user's personal key or private key is denoted by *P*(*i*) that uniquely determines the owner and allows him to decrypt the encrypted broadcast. The broadcast message consists of an enabling block *E* and a cipher block *Y*. The cipher block is the encryption of the actual plaintext data $\bar{m}$ using a secret key '*a*'. That is, $Y = \xi_a(\bar{m})$, where $\xi()$ is the symmetric encryption function for some cryptosystem. The enabling block consists the shares of *a*, which are encrypted using some or all of the *v* keys in the base set *T*. The decryption of enabling block will allow the recovery of the secret key *a*. Every authorized user should be able to recover *a* using his personal key and then decrypt the cipher block using *a* to obtain the plaintext data i.e., $\bar{m} = \zeta_a(Y)$, where $\zeta()$ is the decryption function for the cryptosystem. A collusion of users may conspire and give an unauthorized user a pirate decoder $F$. This pirate decoder will consist of a subset of base keys such that $F \subseteq \bigcup_{i \in W} P(i)$, where $W$ is the coalition of traitors. Once a pirate decoder is found, the broadcaster can trace those who have participated in producing the pirate decoder. Traitor detection will be done by computing $|F \cap P(U)|$ for all users $U$. If $|F \cap P(U)| \geq |F \cap P(V)|$ for all users $V \neq U$, then $U$ is defined to be exposed user.

**Definition 2.2.1** An encryption scheme $c\text{-}TS(k,b,v)$ is called a $c$-Traceability scheme, whenever a pirate decoder produced by a coalition $C$ and $|C| \leq c$, the exposed user $U$ is a member of the coalition.

Suppose $X$ is a set and $B$ is a family of $k$-subsets of $X$ where each $k$-subset is called a block. A Traceability scheme $c\text{-}TS(k, b, v)$ can be considered as a set system $(X, B)$ where each block corresponds to a decoder with $k$ keys from the key set $X$ with the following property.

**Theorem 2.2.1** There exists a $c\text{-}TS(k, b, v)$ if and only if there exists a set system $(X, B)$ such that $|X|=v$, $|B|=b$ and $|P|=k$ for every $P \in B$, with the property that for every choice $d \leq c$ blocks $B_1, ..., B_d \in B$ and for any $k$-subset $F \subseteq \bigcup_{j=1}^{d} B_j$, there does not exist a block $P \in B - \{B_1, ..., B_d\}$ such that $|F \cap B_i| > |F \cap P|$ for $1 \leq j \leq d$. ([see Stinson and Wei 98])

**Definition 2.2.2** A $t$-$(v, k, \lambda)$ design [Colbourn, Dinitz (96)] is a set system $(X, B)$, where $|X|=v$, $|B_i|=k$ for every $B_i \in B$, and every $t$-subset of $X$ occurs in exactly $\lambda$ blocks in $B$.

**2.3 Related work and known bounds**

The construction of frameproof codes from Traceability schemes is presented in [Stinson and Wei 98]. In [Safavi-Naini and Wang 01] lower bounds on maximum number of codewords for frameproof codes and Traceability schemes are presented. Gafni et al [Gafni, Staddon, Yin 99] presented a method for adding any desired level of broadcasting capability to any Traceability scheme and vice versa.

With regard to secure codes Boneh et al [Boneh and Shaw 98] presented an explicit construction for collusion secure codes. Their code has a length of $O\left(c^3 \log\left(1/\varepsilon\right)\right)$ and attains security against coalitions of size $c$ with ε error. Peikert et al [Peikert, Shelat, Smith 03] demonstrated (on lower bound of code length) that no secure code can have a length of $o\left(c^2 \log\left(1/\varepsilon\right)\right)$. Guth and Pitzmann [Guth, Pfitzmann 03] constructed binary $c$-secure codes with $\varepsilon$ error under weak marking assumption, which assumes that adversary can create only a certain percentage of erasures in place of embedded marks in pirate fingerprints and these erasures are (approximately) randomly distributed. In [Trappe, Wu, Wang and Liu 02] a new class of codes, called Anti-Collusion codes (ACC) are proposed which have the property that the composition of any subset of $c$ or fewer code-vectors is unique. Using this property, it is possible to identify groups of $c$ or fewer colluders. Their work presents a construction of binary-valued ACC under the logical AND operation that uses the theory of combinatorial designs ($t$-$(v,k,1)$ designs).

For erasures in non-binary codes Safavi-Naini et al [Safavi-Naini, Wang 03] gave a construction of $q$-ary $c$-secure code that can recover the deleted marks of a shortened fingerprint. In [Safavi-Naini, Wang 03] the codeword is repeated adequate number of times so that at least one copy of the embedded codeword can be recovered and thereby increasing erasure tolerance.

Gossip code is introduced as IPP code in [Lindkvist 01] for collusion secure fingerprinting applications. The Gossip code with $c = q-1$ has been studied in [Lindkvist, Löfvenberg, and M. Svanström 02]. The conditions for a Gossip code to become Traceability (TA) code are presented in [Lindkvist, Löfvenberg, and M. Svanström 02]. For $c = q-1$ the code $G(M,q)$ is a constant weight code whose weight is equal to $\binom{M-1}{q-2}$. For $q = 3$ the use of this code as an error correcting code was analysed in [Svanström 99]. Fernández and Soriano [Fernández, Soriano 02] presented concatenated fingerprinting codes with efficient identification. In this case the inner code use efficient decoding algorithms (Chase algorithms) that correct beyond error correction bound of the code.

## 3   Construction

We present the methods to construct Gossip codes that attain shortest possible length from Traceability schemes and $t$-designs. A program DISCRETA [Betten, Laue, Wassermann 97] to compute $t$-designs is available in public domain. In Gossip codes that achieve the bound, the gossip columns accuse distinct groups and thus every collusion group is an accusation group for some gossip column or the other.

**Lemma 3.1**  A $c$-$Gossip(l, M, q)$ code is shortest if and only if each gossip column accuses $\binom{q-1}{c}$ distinct collusion groups ($c$-groups). (see Lindkvist 01])

The condition for gossip code to achieve its equality bound of code length is that each gossip column should accuse distinct groups. In order to have shortest length each gossip column should accuse maximum possible $\binom{q-1}{c}$ groups that are distinct.

### 3.1 Gossip codes from Partition method and t-designs

Gossip codes that achieve the bound constitute a partition on $\hat{S}$ the set of all user groups (collusion groups of size $c$). This means there exists an equivalence relation $R$, which partitions $\hat{S}$ into equivalence classes. Let $C_i$ and $C_j$ be two collusion sets of size $c$. We say $C_i \, R \, C_j$ ($C_i$ is related to $C_j$) if they are accusations groups for the same gossip column.

Let $X$ be a set of $M$ users. Let $\hat{S}$ be the set of all subsets of $X$ of size $c$. Let $C_i$ and $C_j$ be two members of $\hat{S}$. Consider a $t\text{-}(v,k,\lambda)$ design with $t = c$, $v = M$, $k = q-1$ and $\lambda = 1$. We define a relation $R$ as follows.

$C_i \, R \, C_j$ if and only if $C_i$ and $C_j$ are subsets of the same block in the $t\text{-}(v, k, \lambda)$ design.

It is transparent that the relation $R$ is reflexive, symmetric and transitive and hence an equivalence relation that partitions $\hat{S}$ into equivalence classes.

The construction of $c\text{-}Gossip(l, M, q)$ code from partition method is as follows.

Let $C_1, C_2, ..., C_r$ be the exhaustive members of the $i^{\text{th}}$ equivalence class. We now compute the set $\bigcup_{i=1}^{r} C_i$, which gives the non-zero positions (*column key*) for the $i^{\text{th}}$ gossip column. Repeating the same process for all equivalence classes, one can arrive at all of the *column keys*. From these column keys the indices of non-zero positions for each gossip column are known, and hence a $c\text{-}Gossip(l, M, q)$ code can be constructed.

**Theorem 3.1.1** A $c\text{-}Gossip(l, M, q)$ code exists if and only if $c\text{-}(M, q-1, 1)$ design exists where $l = \binom{M}{c} \Big/ \binom{q-1}{c}$.

**Proof.** Consider a $t\text{-}(v, k, \lambda)$ design with $t = c$, $v = M$, $k = q-1$ and $\lambda = 1$. Then there exists a set system $(X, B)$ where $|X| = v$, $|B_i| = k$ for every $B_i \in B$, and every $c$-subset of $X$ occurs in exactly one block in $B$. Let $B_i$ be the $i^{\text{th}}$ column key of the Gossip code. Then the accusation groups for the $i^{\text{th}}$ gossip column are identical to the $c$-subsets of $B_i$. Since every $c$-subset of $X$ occurs in exactly one block, the accusation groups of all gossip columns are distinct. It is known that number of blocks in $c\text{-}(M, q-1, 1)$ design is $\binom{M}{c} \Big/ \binom{q-1}{c}$. Thus the $t$-design $c\text{-}(M, q-1, 1)$ completely defines $c\text{-}Gossip(l, M, q)$ code. The converse of the theorem is straightforward. □

Lemma 3.1.1 presents the condition for existence of 2-Gossip codes for alphabet sizes 4, 5 and 6. Lemma 3.1.2 presents the existence of a Gossip code with collusion size 3. These lemmas are based on Theorem 3.1.1.

**Lemma 3.1.1** For $3 \leq k \leq 5$, a 2-$Gossip(l, M, k+1)$ code exists if and only if $M \equiv 1$ or $k \mod(k^2 - k)$.

**Proof.** We have shown in *Theorem 3.1.1* that $c$-$Gossip(l, M, q)$ code exists if and only if $c$-$(M, q-1, 1)$ design exists. Thus 2-$Gossip(l, M, k+1)$ code exits if and only if 2-$(M, k, 1)$ design exists. It is known that if $3 \leq k \leq 5$, 2-$(M, k, 1)$ design exists if and only if $M \equiv 1$ or $k \mod(k^2 - k)$ (see Chapter I.2 in [Colbourn, Dinitz (96)]). □

**Lemma 3.1.2** There exists 3-$Gossip(9 \times 82, 82, 11)$ code.

**Proof.** It is known from [Colbourn, Dinitz (96)] that 3-$(p^2 + 1, p + 1, 1)$ design exist when $p$ is prime power. Let $p = 9$ then it follows that 3-$(82, 10, 1)$ design exists. Using *Theorem 3.1.1* construct $c$-Gossip code from 3-$(82, 10, 1)$ design.

We have $v = 82$, $c = 3$, $q = k+1 = 11$

$$\binom{v}{c} = \frac{82 \times 81 \times 80}{6}; \binom{q-1}{c} = \frac{10 \times 9 \times 8}{6}$$

$$\therefore l = 9 \times 82$$

It follows 3-$Gossip(9 \times 82, 82, 11)$ code exists. □

It may be noted that for $c = q-1$, $c$-$Gossip\left(\binom{M}{c}, M, q\right)$ code exists for all values of $M$. For a chosen $M$, $q$ and $c$ the existence a shortest Gossip code is according to *Theorem 3.1.1*. However for any chosen values of $M$, $q$ and $c$, there exists a Gossip code such that $c < q < M$, which may not achieve the shortest length.

The following are two known results [Colbourn, Dinitz (96)] about *t*-designs, which are related to this work.

**Theorem 3.1.2** If $(X, B)$ is a $t$-$(v, k, \lambda)$ design and $S$ is any $s$-element subset of $X$, with $0 \leq s < t$, then the number of blocks containing $S$ is $\lambda_s = |\{P \in B : S \subseteq P\}| = \lambda \cdot \binom{v-s}{t-s} / \binom{k-s}{t-s}$.

**Theorem 3.1.3** If $(X, B)$ is a $t$-$(v, k, \lambda)$ design and $S$ is any $s$-element subset of $X$, with $0 \leq s \leq t$, then the number of blocks does not contain any point of $S$ is $\bar{\lambda}_s = |\{P \in B : P \cap S = \phi\}| = \lambda \cdot \binom{v-s}{k} / \binom{v-t}{k-t}$.

In Theorem 3.1.3 we present some of the properties of shortest Gossip codes.

**Theorem 3.1.4** For a shortest $c\text{-}Gossip(l, M, q)$ code

Length of the code $l = n(M, q) = \binom{M}{c} \big/ \binom{q-1}{c}$

Weight of the code $w(M, q) = \lambda_1 = \binom{M-1}{c-1} \big/ \binom{q-2}{c-1}$

Distance of the code $d(M, q) = l - \binom{M-2}{q-1} \big/ \binom{M-c}{q-1-c}$

**Proof.** From a gossip column of a Gossip code, where $c < q < M$, exactly $\binom{q-1}{c}$ collusion groups are traceable. If $M$ is the total number of codewords in a Gossip code, then the total number of possible pirate groups of size $c$ is given by $\binom{M}{c}$. For $c \leq q-1$, each gossip column at the most can accuse $\binom{q-1}{c}$ groups since there are $(q-1)$ non-zero symbols, which are distinct. In a Gossip code the total number of gossip columns $l \geq \binom{M}{c} \big/ \binom{q-1}{c}$ for $c < q < M$. This gives a lower bound on length of the Gossip code. Hence the shortest Gossip code length attains the equality condition.

We have shown that (Theorem 3.1.1) a shortest $c\text{-}Gossip(l, M, q)$ code exists if and only if $t\text{-}(v, k, 1)$ design exists with $t = c, v = M, k = q-1$ exists. Looking at the $t$-design's blocks we can tell which group's pirate words will have 0 at a given position, and which groups will not. Let $W$ be the collusion group involved in piracy. We shortly write $W = \{i, j\}$ if $w^i$ and $w^j$ are members of $W$. Consider $|W| = 1$, thus from Theorem 3.1.2 the number of blocks that contain $W$ are $\lambda_1$ where $\lambda_1$ is computed with $t = c, v = M$, and $k = q-1$ from Theorem 3.1.2. Thus $\lambda_1$ is the number positions not equal to 0 in any codeword of a Gossip code.

Distance between two codewords is the number of positions in which they differ. If the distance between any two codewords is constant, then distance of the code will be equal to that. Consider $|W| = 2$. If $W \subseteq X - B_i$ (implies $W \cap B_i = \phi$) and $|W| \leq t$ then the collusion group $W$ cannot detect the $i^{th}$ position since the $i^{th}$ position is *zero* in all the codewords of $W$. From Theorem 3.1.3 the number of blocks that do not contain $W$ are $\overline{\lambda_2}$. This means the number of undetectable positions (zeros) for $W$ is $\overline{\lambda_2}$. Thus the number of undetectable positions for any collusition of size '2' is $\overline{\lambda_2}$. Thus the distance between any two code words is $l - \overline{\lambda_2}$ where $\overline{\lambda_2}$ is as defined above with $t = c, v = M, k = q-1$. So the distance of the code becomes $l - \binom{M-2}{q-1} / \binom{M-c}{q-1-c}$.  □

### 3.2 Gossip codes from Traceability Schemes

In this section, we present the construction of shortest length Gossip codes from Traceability Schemes $c\text{-}TS(k, b, v)$ and $t\text{-}(v, k, 1)$ designs. The required condition for construction of Gossip codes from a $c$-Traceability scheme is that the number of private keys in the $c$-Traceability Scheme should be $b = \binom{v}{c} / \binom{k}{c}$. Stinson et al [Stinson and Wei 98] proved the existence infinite class of Traceability Schemes used in our case.

Theorem 3.2.1 presents the construction of Gossip codes from Traceability Schemes. Lemma 3.2.1 presents a condition on $c$-Traceability schemes to be used in construction of Gossip codes.

**Lemma 3.2.1** Let $(X, B)$ denote the set system corresponding to $c\text{-}TS(k, b, v)$ with $B_i$ s representing the private keys of $c\text{-}TS(k, b, v)$ and $c\text{-}Gossip(b, v, k+1)$ denote the Gossip code constructed from $c\text{-}TS(k, b, v)$. If $|B_i \cap B_j| \geq c$ for $i \neq j$, then the accusation groups of the gossip columns are not distinct.

**Proof.** Assume $|B_i \cap B_j| \geq c$. Then there exist at least $c$ elements say $\{x_1,...,x_c\}$ which are common in both $B_i$ and $B_j$. Since the $c$-group $\{x_1,...,x_c\}$ is formed by both $B_i$ and $B_j$, the $c$-groups formed by $B_i$ s are not distinct. Further, the set $\{x_1,...,x_c\}$ will be accusation set for both $i^{\text{th}}$ and $j^{\text{th}}$ gossip columns. Hence the accusation sets corresponding to gossip columns in the Gossip code are not distinct. □

**Theorem 3.2.1** If $c\text{-}TS\left(k, \binom{v}{c}/\binom{k}{c}, v\right)$ exists then $c\text{-}Gossip\left(\binom{v}{c}/\binom{k}{c}, v, k+1\right)$ code exists.

**Proof.** For a $c\text{-}TS(k, b, v)$, the total number of $c$-groups that can be formed from $v$ users is $\binom{v}{c}$. Then the maximum number of $c$-groups that can be formed by each private key containing $k$ elements is $\binom{k}{c}$. If the maximum number of $c$-traceable keys is equal to $\binom{v}{c}/\binom{k}{c}$, then $c$-groups formed by each private key should be distinct. It follows that $|B_i \cap B_j| < c$ for $i \neq j$ where $B_i$ s correspond to $(X, B)$ representation of a $c\text{-}TS(k, b, v)$. Let each $B_i$ be equal to one column key in the Gossip code. Then the $c$-groups created by $k$ elements of each $B_i$ i.e., $\binom{k}{c}$ groups will be the accusation groups for the $i^{th}$ gossip column. Since $|B_i \cap B_j| < c$, from Lemma 3.2.1 each column in Gossip code accuses $\binom{k}{c}$ distinct groups. If we consider $v$ codewords in the Gossip code, equal to the number of base keys in $c\text{-}TS(k, b, v)$, then the code is a $c\text{-}Gossip\left(\binom{v}{c}/\binom{k}{c}, v, k+1\right)$ code, and each column contains $k+1$ distinct elements $(0 \text{ to } k)$. Since we have $l = \binom{v}{c}/\binom{k}{c}$, the Gossip code achieves the minimum code length. □

The following theorem presents the converse of Theorem 3.2.1, i.e., the construction of Traceability Schemes from Gossip codes.

**Theorem 3.2.2** If $c\text{-}Gossip\left(\binom{M}{c}/\binom{q-1}{c}, M, q\right)$ code exists then $w\text{-}TS\left(q-1, \binom{M}{c}/\binom{q-1}{c}, M\right)$ exists where $w = \left\lfloor \sqrt{(q-2)/(c-1)} \right\rfloor$.

**Proof.** Let $(X, B)$ be the set representation of the Gossip code. Then we have $|X| = M$, and $|B_i| = q-1$ for every $B_i \in B$, where each block $B_i$ correspond to one column key. The number of blocks in $B$ is equal to $\binom{M}{c}/\binom{q-1}{c}$. Since the Gossip code achieves the minimum code length the accusation sets of each gossip column are distinct. This implies that this $(X, B)$ system represent a $t$-$(v, k, \lambda)$ design, where $t = c, v = M, k = q-1,$ and $\lambda = 1$. This implies there exists corresponding $w\text{-}TS\left(q-1, \binom{M}{c}/\binom{q-1}{c}, M\right)$ (see Theorem 3.2 in [Stinson and Wei 98]) where $w = \left\lfloor \sqrt{(k-1)/(c-1)} \right\rfloor$. □

**Lemma 3.2.2** There exists a $2\text{-}Gossip(v(v-1)/20, v, 6)$ code for all $v \equiv 1 \text{ or } 5 \mod 20$.

**Proof.** Stinson and Wei (see Theorem 3.4 in [Stinson and Wei 98] showed that there exists $2-TS(5, v(v-1)/20, v)$ whenever $v \equiv 1 \text{ or } 5 \mod 20$. Observe that $\binom{v}{2}/\binom{5}{2} = v(v-1)/20$. From Theorem 3.2.1 it follows that $2\text{-}Gossip(v(v-1)/20,\ v,\ 6)$ code exists. □

### 3.3 Embedded Gossip codes

In many cases, the number of users in a scheme will increase after the system is set up. Initially, the data supplier will construct a scheme that will accommodate a fixed number of users say $M$. If the number of users exceeds $M$, we need to extend the scheme that is compatible with the existing scheme. Such scenarios are possible in Traceability Schemes apart from digital fingerprinting applications. In Traceability Schemes, it is not advisable to change the keys already issued, when the users in the system grow. Embedded Traceability schemes extend the scheme compatible to the existing Traceability Scheme. Embedded Gossip codes (see *Definition 3.3.1*) can be used to construct these Embedded Traceability schemes apart from fingerprinting codes. The construction of embedded Gossip codes is as follows.

**Definition 3.3.1** Let $\Gamma$ be a $c\text{-}Gossip(l,\ M,\ q)$ code and $\Gamma'$ be a $c-Gossip(l',\ M',\ q)$ code, where $M < M'$ and $l < l'$. Suppose that for every codeword $x \in \Gamma$ there exists a codeword $x' \in \Gamma'$ such that the first $l$ symbols are the same as that of $x$. Then we say $\Gamma$ is embedded into $\Gamma'$.

It is simpler to understand the embedding in terms of set systems. Let $(X,\ B)$ and $(X',\ B')$ be two set systems. $(X,\ B)$ is said to be embedded into $(X',\ B')$ if $X \subseteq X'$ and $B \subseteq B'$. Suppose $(X,\ B)$ correspond to $t\text{-}(v,\ k,\ \lambda)$ and $(X',\ B')$ to $t\text{-}(v',\ k,\ \lambda)$ design. Then $t\text{-}(v,\ k,\ \lambda)$ is embedded into $t\text{-}(v',\ k,\ \lambda)$ if and only if $(X,\ B)$ is embedded into $(X',\ B')$.

**Lemma 3.3.1** A $c\text{-}Gossip(l,\ M,\ q)$ code can be embedded into $c\text{-}Gossip(l',\ M',\ q)$ code if and only if the respective $c\text{-}(M,\ q-1,\ 1)$ design is embedded in $c\text{-}(M',\ q-1,\ 1)$.

**Proof.** The proof follows from *Theorem 3.1.1* since a $c\text{-}Gossip(l,\ M,\ q)$ code is identical to $c\text{-}(M,\ q-1,\ 1)$ design in set system representation. □

As an example $2\text{-}Gossip(7,\ 7,\ 4)$ code can be embedded into $2\text{-}Gossip(35,\ 15,\ 4)$ code.

## 4   Erasures in Gossip codes

We consider erasures in Gossip codes that achieve minimum possible code length. In our discussions, *erasure* means *a non-alphabet symbol* chosen by the pirates in a detected position of the embedded fingerprint. The pirates can *find* only those alphabet symbols that match with any one of their copies at each detected position.

We assume that this marking assumption [Boneh and Shaw 98] holds good in our model. In general, the embedding mechanism, which encodes codewords into digital objects, will determine the type of alphabet symbols or erasures that are possible at each detected position for creating an illegal copy.

Our model assumes that pirates can create one of the alphabet symbols matching any one of their copies, or an erasure in the place of a detected mark. When the pirates choose a symbol that matches with a symbol in the alphabet, it is not be possible to differentiate between the symbols chosen by the pirates and the valid symbols embedded under usual fingerprinting methods. This may lead to errors in tracing. For this reason, the embedding algorithm should be so chosen that it hides the actual alphabet from being detected by the pirates (through some encoding mechanism). We now discuss the *erasures* in Gossip codes.

**Definition 4.1** A *Fingerprinting System with Optional Erasures* (FSOPE) is a system where the pirate controlling a group of traitors may *optionally* choose to create erasures at the detected marks, i.e., the pirate may create an alphabet symbol that he can *find* or a non-alphabet symbol.

*FSOPE* include the following cases:
- *No Erasures*
- *Selective Erasures*
- *Only Erasures*

### 4.1 Gossip codes with no erasures

In this case, the pirates are not allowed to create erasures, however, at each position they are free to choose the alphabet symbols they find in their copies. The tracing algorithm identifies a culprit from the extracted codeword. The tracing mechanism is obvious when the coalition size is equal to one. This is because the pirate's codeword has to match with one of the valid codewords of the Gossip code. We consider e.g. 2.2.1, i.e., *2*-Gossip (7, 7, 4) code to illustrate tracing in Gossip codes with no erasures.

#### 4.1.1 Tracing with no erasures

Let the pirate set be $W = \{w^1, w^2\}$ where $w^i$ is the $i^{th}$ codeword in e.g. 2.1.1. Let the pirate word found in the illegal copy be $x = \{2, 0, 0, 0, 0, 0, 0\}$. Clearly $x \in D(w^1, w^2)$, but $x$ also belongs to $D(w^2, w^3)$ where $D(w^i, w^j)$ is the descendent set of $\{w^i, w^j\}$.

Consider the non-zero value in the pirate word $x$ of the above example. Its position is first. Observe that the second codeword contains 2 in the first position. We accuse the $2^{nd}$ user as culprit because the position of non-zero alphabet symbol and the value in the pirate word match with that of the second user's codeword. It may also be noted that the $2^{nd}$ user is the member of both the coalitions that are capable of creating the pirate word.

These codes can be constructed from *Theorem 3.1.1* and *Theorem 3.2.1*. Please see the Appendix for details.

**Lemma 4.1.1** Every pirate word created from a Gossip code containing at least one non-zero alphabet symbol can reveal one member of the collusion.

**Proof.** The construction of the Gossip code makes it clear that every non-zero symbol in the pirate word can reveal one member of the collusion. Let the $i^{th}$ position in the pirate word be a valid alphabet symbol say '1', then there exists exactly one codeword say $j^{th}$ codeword, which contains '1' in the $i^{th}$ position. It is clear that without the involvement of $j^{th}$ user, it is impossible to create the pirate word $x$ and so the algorithm accuses $j^{th}$ user. □

The general construction of a Gossip code that can trace all active colluders (in *no erasures* case) who contributed to pirate copy is as follows:

Choose the number of codewords equal to the number of alphabet symbols i.e., $M = q$. Let the collusion size $c$ be equal to $q-1$ and the number of gossip columns be $l = q$. Construct $q$ gossip columns, where each column contains the alphabet symbols $\{0,...,q-1\}$ only once. It is known from [Lindkvist 01] that when $c = q-1$, there exists a Gossip code such that each column accuses one of the collusion sets.

If $M = q$ then $c$-$Gossip(1, M, M)$ code is shortest code since all symbols are distinct. But for concatenated codes we consider inner code with $l = q = M$.

**Lemma 4.1.2.** A $(q$-$1)$-$Gossip(q, q, q)$ code has a deterministic tracing algorithm and the tracing algorithm accuses the active traitors.

**Proof.** In a $(q$-$1)$-$Gossip(q, q, q)$ code every symbol in the alphabet appears only once in each column. So the alphabet symbol 0 also contributes to tracing and each gossip column can trace $\binom{q}{c}$ pirate groups. This leads to tracing of all the active users who contributed in creating the pirate word by using their legitimate copies.

Example 4.1.1  4-Gossip (5, 5, 5) code.

| 0 | 1 | 1 | 1 | 1 |
|---|---|---|---|---|
| 1 | 2 | 2 | 2 | 0 |
| 2 | 3 | 3 | 0 | 2 |
| 3 | 0 | 4 | 3 | 3 |
| 4 | 4 | 0 | 4 | 4 |

**4.2   Gossip codes with selective erasures**

In this case users are free to create erasures or the alphabet symbols they *found* in their copies at each detected position of the collusion set. When a pirate copy is created by a colluded set the undetected positions of the collusion remain the same in the copy.

### 4.2.1 Tracing with selective erasures

*Lemma 4.1.1* holds good for all Gossip codes even in presence of erasures, thus any pirate word containing one non-zero symbol can reveal one traitor. In a *c-Gossip*($l$, $M$, $q$) code if the pirates create only erasures at all detected positions then the tracing is same as in *only erasures* case.

**Lemma 4.2.1** The ($q$-1)-*Gossip*($q$, $q$, $q$) code can trace all the pirate words except ($e,e,e,e,e... l$ times) to reveal at least one culprit.

**Proof.** In *selective erasures* case of a ($q$-1)-*Gossip*($q$, $q$, $q$) code every alphabet symbol (including 0) can identify a culprit. But if all the symbols are erasures nobody can be accused.

In e.g. 2.2.1 it can be seen that the pirate word ($e,e,e,e,e$) contains all erasures and so the tracing will not yield any pirate. It will be the same as choosing only erasures in detected positions, since all the positions are detected for a collusion of size $c = 2$.

### 4.3 Gossip codes with only erasures

An embedded position is detected when its value is found different in two copies during comparison. In *only erasures* case, the pirates are allowed to create only erasures in the detected positions and cannot choose any valid alphabet symbol. In *only erasures* case (of any shortest Gossip code) it is always true that each pirate group can create only one pirate word. Let *c-Gossip*($l$, $M$, $q$) code be a code with $c = q-1$ and $M = 2.(q-1)$. Then the number of zeros in each gossip column is equal to the number of pirates $c$, i.e., $c = M - (q-1)$. This will guarantee a unique coordinate (in the pirate word) for each pirate group, which has only zeros. Thus the pirate words created by different pirate groups are all distinct. Here is an example.

Example 4.3.1  2-Gossip(6, 4, 3) code

| 1 | 1 | 0 | 0 | 1 | 0 |
|---|---|---|---|---|---|
| 2 | 0 | 1 | 1 | 0 | 0 |
| 0 | 2 | 2 | 0 | 0 | 1 |
| 0 | 0 | 0 | 2 | 2 | 2 |

Here, the number of pirate sets of size 2 are equal to $\binom{4}{2}$ =6. The exhaustive list of the pirate words (equal to number of pirate groups here) created by all the collusion sets are as follows, where *e* denotes an erasure at all detected positions and *0* means the positions are undetected.

| S.No | Pirate Sets | Descendent sets of Pirates |
|---|---|---|
| 1 | {1, 2} | (e, e, e, e, e, 0) |
| 2 | {1, 3} | (e, e, e, 0, e, e) |
| 3 | {1, 4} | (e, e, 0, e, e, e) |
| 4 | {2, 3} | (e, e, e, e, 0, e) |
| 5 | {2, 4} | (e, 0, e, e, e, e) |
| 6 | {3, 4} | (0, e, e, e, e, e) |

*Table 1: Collusion sets vs. pirate fingerprints*

### 4.3.1 Tracing with only erasures

We observe that each collusion set creates a unique fingerprint in o*nly erasures* case, i.e., the number of pirate fingerprints is equal to number of collusion sets. Moreover all the pirate words are distinct. Since there is one-to-one matching between a particular collusion set and the pirate fingerprint, it is possible to find the collusion group as a whole. This tracing method has $O\binom{M}{c}$ complexity where $c$ is collusion size. An alternative method is to find all the zeros in the pirate word, and accuse everybody who has zeros in *all* these positions. This works even if the actual coalition size is less than $c$, and runs in time $O(M \cdot l)$ where $M$ is the number of users and $l$ is the length of the code. This is significantly more efficient than $O\binom{M}{c}$, which is the running time of the earlier algorithm.

### 4.3.2 General case for only erasures

Let $(X, B)$ denote the equivalent set system of a $t$-design that corresponds to a $c$-$Gossip(l, M, q)$ code. We shortly write $W = \{i, j\}$ if $w^i$ and $w^i$ are members of the collusion group $W$. Looking at the $t$-design's blocks we can tell which group's pirate words will have 0 at a given position, and which groups will have erasures. For example the block $B_i$ gives the list of pirate groups, which will not be able to detect the $i^{th}$ position. If $W \subseteq X - B_i$ (implies $W \cap B_i = \phi$) and $|W| \le t$ then the collusion group $W$ cannot detect the $i^{th}$ position. $W$ will create an erasure (since the mark is detected) in $j^{th}$ position if $W \cap B_j \ne \phi$. Let $x$ be the pirate word created by a collusion group $W$ and $x_i$ be the $i^{th}$ position in the pirate word. The pirate word created by $W$ can be completely specified by the blocks of the $t$-design.

   $x_j$ is zero (undetected position) if $W \cap B_j = \phi$ and
   $x_j$ is $e$ (detected) if $W \cap B_j \ne \phi$

From the above discussion and Theorem 3.1.2 can be seen that the number of non-zero symbols in each codeword of a *c-Gossip*($l$, $M$, $q$) code is equal to $\lambda_1$. Thus each codeword contain $l - \lambda_1$ zeros.

**Lemma 4.3.2.1** If *c-Gossip*($l$, $M$, $q$) code is a Gossip code with minimum code length, then the number of undetectable positions (number of zeros in only erasures case) for a collusion $W$ of size $d \leq c$ in the pirate word is $l - \left( \sum_{i=1}^{d} (-1)^{i-1} \binom{d}{i} \lambda_i \right)$ where $\lambda_i$ is defined as in Theorem 3.1.2 above, with $\lambda = 1$, $v = M$ and $t = c$.

**Proof.** Let $W$ be the collusion set of size $d \leq c$. The number of blocks in the $t$-($v$, $k$, $\lambda$) design disjoint from $W$, i.e., total number of blocks excluding the blocks that contain any member of $W$, is equal to the number of undetectable positions for the collusion. The proof follows from *Theorem 3.1.1* and the above discussion. □

For example consider 2-Gossip (7, 7, 4) code presented in e.g. 2.2.1. The *t*-design corresponding to this Gossip code is 2-(7, 3, 1) design, which is given by

$X = \{1, 2, 3, 4, 5, 6, 7\}$
$B_1 = \{1, 2, 3\}, B_2 = \{1, 4, 5\}, B_3 = \{1, 6, 7\}, B_4 = \{2, 4, 6\},$
$B_5 = \{2, 5, 7\}, B_6 = \{3, 5, 6\}, B_7 = \{3, 4, 7\}$

Let $W$ be a collusion set, and let $W = \{1, 2\}$.

Number of blocks that do not contain any member of $W$ = Total number of blocks – Number of blocks that contain 1 – Number of blocks that contain 2 + Number of blocks that contain $\{1, 2\}$ = $l - 2.\lambda_1 + 1$ = 7-(2×3)+1=2

This implies in only erasures case for 2-Gossip (7, 7, 4) code each pirate word contain two zeros. In shortest Gossip code, if $c < M - (q-1)$ then the number of zeros in a column is more than $c$. This ensures that there is at least one undetected position for the pirate words. It is known that $2-(p^2 + p + 1, p + 1, 1)$ design exists when $p$ is prime power. When a Gossip code is constructed from $2-(p^2 + p + 1, p + 1, 1)$ design, the number of zeros in each Gossip column $= M - (q-1) = p^2 + p + 1 - (p+1) = p^2 > c$. Also the number of $c$-groups for which, a given position in the pirate code word is undetectable is $\binom{p^2}{c}$. Thus for the code in e.g. 2.1.1, the number of zeros in each column is 4, and $\binom{4}{2} = 6$ pirate words contain 0 in a given position.

Consider the Gossip code as given in e.g. 2.1.1, where $M = 7$, $q = 4$ and $c = 2$. Here, the number of pirate sets of size 2 are equal to $\binom{7}{2}=21$. The exhaustive list of the pirate words (equal to number of pirate groups here) created by all the collusion sets are as follows.

| S.No | Pirate Sets | Descendent sets of Pirates |
|---|---|---|
| 1 | {1, 2} | (e, e, e, e, e, 0, 0) |
| 2 | {1, 3} | (e, e, e, 0, 0, e, e) |
| 3 | {1, 4} | (e, e, e, e, 0, 0, e) |
| 4 | {1, 5} | (e, e, e, 0, e, e, 0) |
| 5 | {1, 6} | (e, e, e, e, 0, e, 0) |
| 6 | {1, 7} | (e, e, e, 0, e, 0, e) |
| 7 | {2, 3} | (e, 0, 0, e, e, e, e) |
| 8 | {2, 4} | (e, e, 0, e, e, 0, e) |
| 9 | {2, 5} | (e, e, 0, e, e, e, 0) |
| 10 | {2, 6} | (e, 0, e, e, e, e, 0) |
| 11 | {2, 7} | (e, 0, e, e, e, 0, e) |
| 12 | {3, 4} | (e, e, 0, e, 0, e, e) |
| 13 | {3, 5} | (e, e, 0, 0, e, e, e) |
| 14 | {3, 6} | (e, 0, e, e, 0, e, e) |
| 15 | {3, 7} | (e, 0, e, 0, e, e, e) |
| 16 | {4, 5} | (0, e, 0, e, e, e, e) |
| 17 | {4, 6} | (0, e, e, e, 0, e, e) |
| 18 | {4, 7} | (0, e, e, e, e, 0, e) |
| 19 | {5, 6} | (0, e, e, e, e, e, 0) |
| 20 | {5, 7} | (0, e, e, 0, e, e, e) |
| 21 | {6, 7} | (0, 0, e, e, e, e, e) |

*Table 2: Collusion sets vs. pirate fingerprints (2-Gossip(7, 7, 4) code)*

**Theorem 4.3.2.1** If $c$-$Gossip(l, M, q)$ code is a Gossip code with shortest length such that $c < M - (q-1)$, then no two collusion groups can create the same pirate word in fingerprinting with only erasures.

**Proof.** In *only erasures* case (of any Gossip code) it is always true that each pirate group can create only one pirate word. If $c < M - (q-1)$ in a Gossip code with minimum possible length the pattern of undetectable zeros is different for each group. Thus the pirate words are unique. If two pirate words can create same pirate word then each gossip column cannot accuse $\binom{q-1}{c}$ distinct groups, which is a contradiction. □

Examples for these codes are Gossip codes constructed from 2-$(p^2 + p +1, p+1, 1)$ and 3-$(p^2 +1, p+1, 1)$ designs where $p$ is prime power and greater than 2. Specific examples are 3-*Gossip*(30, 10, 5), and 2-*Gossip*(21, 21, 6) codes apart from 2-*Gossip*(7, 7, 4) code.

## 5 Concatenated Codes

A Concatenated code consists of an inner code and an outer code, where the symbols in the outer code are the codewords of the inner code. Thus, every codeword in the Concatenated code is the concatenated collection of inner codewords. Further, the number of codewords in inner code is equal to the number of alphabet symbols in the outer code. We denote the codewords of the inner code as $\{\overline{0},...,\overline{q-1}\}$ for convenience. We replace the symbols of outer code with inner codewords, for constructing the concatenated code.

When a frameproof code is used as inner code, the pirate members cannot create alphabet symbols not found in their copies in the detected positions of the concatenated code. Thus, whenever a mark is detected during comparison in the concatenated code, the pirates need to create an alphabet symbol they find in their copies, or a non-alphabet symbol in the detected positions. Thus this concatenated code presents a sensible way of implementing fingerprinting schemes under the pirate model assuming marking assumption [Boneh and Shaw 98] for non-binary codes. Consider a frameproof code with $M = 4$, $q = 2$ and $c = 2$. Let $\overline{0}$ denote the first inner codeword, $\overline{1}$ denote the second, and so on.

Example 5.1  2-FP(3, 4, 2) code

| 1 | 0 | 0 |   | $\overline{0}$ |
|---|---|---|---|---|
| 0 | 1 | 0 | ≡ | $\overline{1}$ |
| 0 | 0 | 1 |   | $\overline{2}$ |
| 1 | 1 | 1 |   | $\overline{3}$ |

If we use the Gossip code in e.g. 2.2.1 as outer code for the Concatenated code, the Concatenated code will be:

Example 5.2  2-Gossip(7, 7, 4) concatenated code

| $\overline{1}$ | $\overline{1}$ | $\overline{1}$ | $\overline{0}$ | $\overline{0}$ | $\overline{0}$ | $\overline{0}$ |   | $w^1$ |
|---|---|---|---|---|---|---|---|---|
| $\overline{2}$ | $\overline{0}$ | $\overline{0}$ | $\overline{1}$ | $\overline{1}$ | $\overline{0}$ | $\overline{0}$ |   | $w^2$ |
| $\overline{3}$ | $\overline{0}$ | $\overline{0}$ | $\overline{0}$ | $\overline{0}$ | $\overline{1}$ | $\overline{1}$ |   | : |
| $\overline{0}$ | $\overline{2}$ | $\overline{0}$ | $\overline{2}$ | $\overline{0}$ | $\overline{0}$ | $\overline{2}$ | ≡ | : |
| $\overline{0}$ | $\overline{3}$ | $\overline{0}$ | $\overline{0}$ | $\overline{2}$ | $\overline{2}$ | $\overline{0}$ |   | : |
| $\overline{0}$ | $\overline{0}$ | $\overline{2}$ | $\overline{3}$ | $\overline{0}$ | $\overline{3}$ | $\overline{0}$ |   | : |
| $\overline{0}$ | $\overline{0}$ | $\overline{3}$ | $\overline{0}$ | $\overline{3}$ | $\overline{0}$ | $\overline{3}$ |   | $w^7$ |

The collusion size $c$ is two for the concatenated code in e.g. 5.2. Now, we discuss the tracing for the Concatenated code defined above.

Consider the collusion set $W$, consisting of $w^1$ and $w^2$. The descendent set is $D(W) = \{(a_1,...,a_5, \overline{0}, \overline{0}) / a_1 \in \{\overline{1}, \overline{2}\}, a_i \in \{\overline{1}, \overline{0}\}$ for $2 \leq i \leq 4\}$ since the last two positions are undetected. The choices in the other positions are based on the choices available in the inner code (i.e., code given in e.g 5.1.) Let a pirate word created by the collusion set $W = \{w^1, w^2\}$ be $x = \{0\,0\,1\ 0\,0\,0\ 0\,0\,0\ 0\,1\,0\ 0\,0\,0\ 1\,0\,0\ 1\,0\,0\}$. This can be re-written as $\{\overline{2}, \overline{e}, \overline{e}, \overline{1}, \overline{e}, \overline{e}, \overline{0}, \overline{0}\}$. This is selective erasures case of simple Gossip code. Further the pirate word contains a nonzero symbol. In this case the tracing algorithm runs in time $O(M \cdot l)$ and will trace $w^2$.

**5.1    Concatenated Gossip codes with Selective erasures**

In this construction, the outer code is also a Gossip code (that has a tracing algorithm) and the inner code has a deterministic decoding algorithm, so the Concatenated code also traces back the members of the pirate group. Dealing with erasures in inner Gossip codes would automatically mean dealing with erasures in Concatenated codes. Consider a Gossip code with $M = q = 4$, and $c = 3$. It would result in the following code, where $\overline{0}$ denotes the first inner codeword, $\overline{1}$ denote the second, and so on.

Example 5.1.1  3-Gossip(4, 4, 4) code

| 1 | 1 | 1 | 0 |     | $\overline{0}$ |
|---|---|---|---|-----|---|
| 2 | 2 | 0 | 1 | $\equiv$ | $\overline{1}$ |
| 3 | 0 | 2 | 2 |     | $\overline{2}$ |
| 0 | 3 | 3 | 3 |     | $\overline{3}$ |

This code can trace all its pirates if the collusion size is less than or equal to three (refer *Lemma 4.1.2*). If we use the Gossip code in e.g. 2.1.1 as outer code for the Concatenated code, the Concatenated code will be:

Example 5.1.2  2–Gossip(7, 7, 4) concatenated code

| $\overline{1}$ | $\overline{1}$ | $\overline{1}$ | $\overline{0}$ | $\overline{0}$ | $\overline{0}$ | $\overline{0}$ |     | $w^1$ |
|---|---|---|---|---|---|---|---|---|
| $\overline{2}$ | $\overline{0}$ | $\overline{0}$ | $\overline{1}$ | $\overline{1}$ | $\overline{0}$ | $\overline{0}$ |     | $w^2$ |
| $\overline{3}$ | $\overline{0}$ | $\overline{0}$ | $\overline{0}$ | $\overline{0}$ | $\overline{1}$ | $\overline{1}$ |     | : |
| $\overline{0}$ | $\overline{2}$ | $\overline{0}$ | $\overline{2}$ | $\overline{0}$ | $\overline{0}$ | $\overline{2}$ | $\equiv$ | : |
| $\overline{0}$ | $\overline{3}$ | $\overline{0}$ | $\overline{0}$ | $\overline{2}$ | $\overline{2}$ | $\overline{0}$ |     | : |
| $\overline{0}$ | $\overline{0}$ | $\overline{2}$ | $\overline{3}$ | $\overline{0}$ | $\overline{3}$ | $\overline{0}$ |     | : |
| $\overline{0}$ | $\overline{0}$ | $\overline{3}$ | $\overline{0}$ | $\overline{3}$ | $\overline{0}$ | $\overline{3}$ |     | $w^7$ |

The collusion size $c$ is two for the concatenated code. Now, we discuss the optional erasures case for the Concatenated code defined above. In this case, the pirates are allowed to create an alphabet symbol they *see* in the detected positions.

### 5.1.1 Tracing

Consider the collusion set $W$, consisting of $w^1$ and $w^2$. The descendent set is given by $D(W) = \{(a_1,...,a_5, \overline{0}, \overline{0}) / a_1 \in \{\overline{1}, \overline{2}\}, a_i \in \{\overline{1}, \overline{0}\}$ for $2 \leq i \leq 4\}$ since the last two positions are undetected. The choices in the other positions are based on the choices available in the inner code (i.e., code given in e.g. 5.1.1)

Let the pirate word created by the collusion set $W$ be $x = \{2\ 2\ 2\ 2\ \ 2\ 2\ 0\ 1\ \ 1\ 1\ 1\ 1\ \ 1\ 1\ 0\ 1\ \ 1\ 2\ 1\ 1\ \ 1\ 1\ 1\ 0\ \ 1\ 1\ 1\ 0\}$. This can be re-written as $\{\overline{e}, \overline{1}, \overline{e}, \overline{0}, \overline{e}, \overline{0}, \overline{0}\}$. This is same as simple Gossip code with selective erasures. This will reveal a pirate namely $w^1$. Suppose the pirateword is $x = \{2\ 2\ 2\ 2\ \ 1\ 2\ 0\ 0\ \ 1\ 1\ 1\ 0\ \ 1\ 1\ 0\ 1\ \ 1\ 2\ 1\ 1\ \ 1\ 1\ 1\ 0\ \ 1\ 1\ 1\ 0\}$. This can be rewritten as $\{\overline{e}, \overline{e}, \overline{e}, \overline{0}, \overline{e}, \overline{0}, \overline{0}\}$. This does not contain any non-zero symbol of the outer code. But here the inner codewords that are exposed by applying tracing algorithm on inner pirate word $\{2\ 2\ 2\ 2\}$ are $\{\overline{1}, \overline{2}\}$. Similarly, all other inner pirate words, reveal $\{\overline{0}, \overline{1}\}$ except the last two pirate words, which reveal $\overline{0}$. With this information, we can completely identify the set of pirates. This traces the pirate set to be $\{w^1, w^2\}$.

**Lemma 5.1:** From an inner $c$-*Gossip*($l$, $M$, $q$)) code and an outer $c$-*Gossip*($l'$, $M'$, $q'$) code is a concatenated IPP code can be constructed provided $M = q'$.

To construct a generic concatenated code, the condition that must be fulfilled is that the number of inner codewords is equal to the number of alphabet symbols of the outer code Further, this concatenated code is an IPP code, since the inner and outer codes are IPP codes. Such a concatenated code can be constructed by considering *e.g* 4.3.1 as inner code and e.g. 2.1.1 as outer code. The *no erasures* and o*nly erasures* cases in these concatenated codes are straightforward and omitted here as the tracing in the inner codewords is similar to Gossip codes for *no erasures and only erasures*.

### 5.2 Performance of Gossip codes

One advantage of Gossip codes is that they can accuse a traitor deterministically. Nevertheless all the previous constructions were probabilistic in some way or other. In general, the focus in all probabilistic code constructions is to keep the length of each codeword short. Short length is important to design efficient and effective embedding techniques of a codeword into an object. Some of the previous probabilistic codes managed to achieve polylog size in the number of the users. But the length of the code tends to be very large (to be interpreted as proportional to the number of users) if deterministic full tracing solutions are sought. Thus Gossip codes in general, have much greater length than probabilistic codes. Our constructions for Gossip codes achieve the minimum possible code length specified for Gossip codes.

The additional advantage of Gossip codes comes from the fact that they can withstand erasures and tracing is possible in presence of erasures. For improved results concatenated codes with outer code as Gossip code and inner codes with probabilistic tracing can be used.

### 5.3    Embedding fingerprints

We present simple watermarking technique, which are based on wavelet transforms to embed (hide) the codewords of a Gossip code into images. The distributor may choose any of the embedding method from the available watermarking techniques based on the requirement and type of the media.

#### 5.3.1    Wavelet watermarking

The Discrete Wavelet Transform (DWT) separates an image into a lower resolution approximation (*LL*) as well as horizontal (*HL*), vertical (*LH*) and diagonal (*HH*) detail components. The process can be repeated to compute multiple 'scale' wavelet decomposition similar to the two-scale wavelet transform as shown below.

One of the advantages of the using wavelet transform is that it is believed to more accurately model the Human Visual System (HVS) as compared to Fast Fourier Transform (FFT) and Discrete Cosine Transform (DCT). This allows us to use higher energy watermarks in regions that the HVS is known to be less sensitive to; such as the high-resolution detail bands *LH*, *HL*, and *HH*.

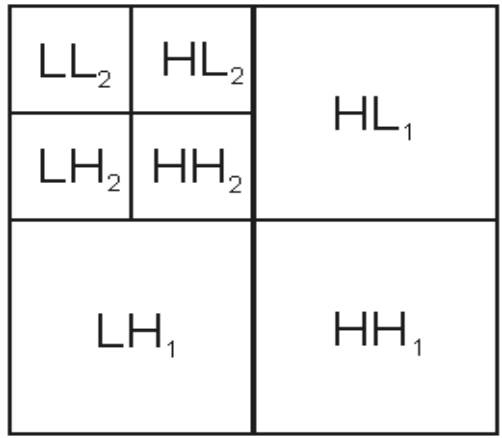

*Figure 1: Wavelet Decomposition*

#### 5.3.2    Algorithm

We use a straightforward technique for embedding the Gossip codeword (figure 2) as a watermark sequence in the detail bands according to the equation shown below:

$$I_{V_{a,b}} = \begin{cases} V_i + \alpha.|V_i|.wm_i, & a,b \in HL, LH \\ V_i & a,b \in LL, HH \end{cases}$$

where $V_i$ denotes the coefficient of the transformed image, $wm_i$ one bit of the watermark to be embedded, and a scaling factor. In order to detect the watermark, we generate the same watermark sequence and determine its correlation with the two transformed detail bands. If the correlation exceeds some threshold $\hat{T}$, the watermark is detected. Thus, in this method watermark detection does not require the original image. This method can be easily extended to embed multiple watermarks into the image. The robustness evaluation (and tracing) was limited to testing against JPEG compression and the addition of random noise.

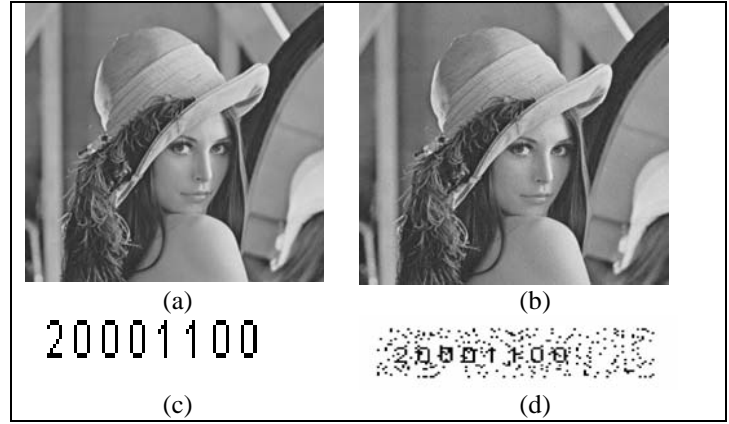

*Figure 2: a) Original Image b) Watermarked Image c) Embedded Watermark d) Recovered Watermark*

## 6 Conclusion

Gossip codes that achieve the shortest code length can be constructed from class $t$-designs and $c$-Traceability Schemes. In this work, two methods of construction for shortest Gossip codes are presented. The converse part i.e., construction of Traceability Schemes and $t$-designs from Gossip codes is also feasible. These results are reliable to determine whether a shortest Gossip code exists or not, for the chosen code parameters $M$, $q$ and $c$. Gossip codes, as IPP codes, can identify one parent of the pirate copy during tracing. They allow deterministic tracing of pirates and also come up with a tracing algorithm, whereas earlier works on fingerprinting are usually probabilistic in some way or the other. The construction of embedded Gossip code is also possible and it can be used for extending an existing Gossip code to a bigger code. These (embedded) codes and Traceability schemes can expand a broadcast application such that it is compatible to the existing scheme. When a frameproof code is used as inner code, the pirate members cannot create alphabet symbols not found in their copies in the detected positions of the concatenated code. Thus, during comparison, whenever a mark is detected in the concatenated code, the pirates need to create an alphabet symbol they find in their copies or a non-alphabet symbol in the detected positions. Thus, this concatenated code presents a sensible way for implementing fingerprinting schemes under pirate model (FSOPE), assuming marking assumption [Boneh and Shaw 98] for non-binary codes. Thus, concatenated codes are presented here for the realization of the pirate model and tracing. There are various advantages of shortest Gossip codes. Shortest Gossip codes are likely to cause less distortion during embedding due to shorter length as compared to normal Gossip codes since there are fewer modifications to be made to the original

**Acknowledgements**

The first author would like to thank Mr. S. Rashmi Dev for suggestions pertaining to this manuscript.

## Appendix

*2-Gossip(21, 21, 6) code from 2-TS (5, 21, 21)*
Base Keys = T = {1 to 21}
Private Key $P(i) = \{2+i,\ 5+i,\ 6+i,\ 11+i,\ 13+i\}$
Total number of pirate groups of size $c = 2$ is equal to

$$\binom{v}{c} = \binom{21}{2} = \frac{21(20)}{2} = 210 \text{ and}$$

$$\binom{q-1}{c} = \binom{k}{c} = \binom{5}{2} = 10\ .$$

So $l = 210/10 = 21$ and $B_1 = P(1)$, ..., $B_i = P(i)$

| User No (*i*) | Private Key P(*i*) |
|---|---|
| 1. | {3, 6, 7, 12, 14} |
| 2. | {4, 7, 8, 13, 15} |
| 3. | {5, 8, 9, 14, 16} |
| 4. | {6, 9, 10, 15, 17} |
| 5. | {7, 10, 11, 16, 18} |
| 6. | {8, 11, 12, 17, 19} |
| 7. | {9, 12, 13, 18, 20} |
| 8. | {10, 13, 14, 19, 21} |
| 9. | {11, 14, 15, 20, 1} |
| 10. | {12, 15, 16, 21, 2} |
| 11. | {13, 16, 17, 1, 3} |
| 12. | {14, 17, 18, 2, 4} |
| 13. | {15, 18, 19, 3, 5} |
| 14. | {16, 19, 20, 4, 6} |
| 15. | {17, 20, 21, 5, 7} |
| 16. | {18, 21, 1, 6, 8} |
| 17. | {19, 1, 2, 7, 9} |
| 18. | {20, 2, 3, 8, 10} |
| 19. | {21, 3, 4, 9, 11} |
| 20. | {1, 4, 5, 10, 12} |
| 21. | {2, 5, 6, 11, 13} |

*Table 3: 2-Traceablity Scheme (2-TS (5, 21,21))*

Considering the first private key $B_1$ we get the following accusation groups. Also each private key, gives the indices of non-zero alphabet symbols, for one gossip column.

| Column | Accusation Groups |
|---|---|
| 1 | {3, 6}, {3, 7}, {3, 12}, {3, 14}, {6, 7}, {6, 12}, {6, 14}, {7, 12}, {7, 14}, {12, 14} |

*Table 4: Accusation groups of 1$^{st}$ column in 2-Gossip (21, 21, 6) code*

Thus the first 10 columns of the Gossip code are:

| 0 | 0 | 0 | 0 | 0 | 0 | 0 | 0 | 5 | 0 |
|---|---|---|---|---|---|---|---|---|---|
| 0 | 0 | 0 | 0 | 0 | 0 | 0 | 0 | 0 | 5 |
| 1 | 0 | 0 | 0 | 0 | 0 | 0 | 0 | 0 | 0 |
| 0 | 1 | 0 | 0 | 0 | 0 | 0 | 0 | 0 | 0 |
| 0 | 0 | 1 | 0 | 0 | 0 | 0 | 0 | 0 | 0 |
| 2 | 0 | 0 | 1 | 0 | 0 | 0 | 0 | 0 | 0 |
| 3 | 2 | 0 | 0 | 1 | 0 | 0 | 0 | 0 | 0 |
| 0 | 3 | 2 | 0 | 0 | 1 | 0 | 0 | 0 | 0 |
| 0 | 0 | 3 | 2 | 0 | 0 | 1 | 0 | 0 | 0 |
| 0 | 0 | 0 | 3 | 2 | 0 | 0 | 1 | 0 | 0 |
| 0 | 0 | 0 | 0 | 3 | 2 | 0 | 0 | 1 | 0 |
| 4 | 0 | 0 | 0 | 0 | 3 | 2 | 0 | 0 | 1 |
| 0 | 4 | 0 | 0 | 0 | 0 | 3 | 2 | 0 | 0 |
| 5 | 0 | 4 | 0 | 0 | 0 | 0 | 3 | 2 | 0 |
| 0 | 5 | 0 | 4 | 0 | 0 | 0 | 0 | 3 | 2 |
| 0 | 0 | 5 | 0 | 4 | 0 | 0 | 0 | 0 | 3 |
| 0 | 0 | 0 | 5 | 0 | 4 | 0 | 0 | 0 | 0 |
| 0 | 0 | 0 | 0 | 5 | 0 | 4 | 0 | 0 | 0 |
| 0 | 0 | 0 | 0 | 0 | 5 | 0 | 4 | 0 | 0 |
| 0 | 0 | 0 | 0 | 0 | 0 | 5 | 0 | 4 | 0 |
| 0 | 0 | 0 | 0 | 0 | 0 | 0 | 5 | 0 | 4 |